\documentclass[twocolumn]{aastex701}

\usepackage{amsmath,amssymb,amsfonts}
\usepackage{graphicx}
\usepackage{bm}
\usepackage{booktabs}

\newcommand{\vect}[1]{\bm{#1}}
\newcommand{\mat}[1]{\mathbf{#1}}
\newcommand{\expect}{\mathbb{E}}
\newcommand{\reals}{\mathbb{R}}

\shorttitle{NestyNet. II. Coherent Function-Space Posteriors}
\shortauthors{Ibata et al.}

\begin{document}

\title{NestyNet. II. Coherent Function-Space Posteriors from Scientific Neural Surrogates\\
(or How to Avoid Expensive MCMC)}

\author[0000-0002-3292-9709]{Rodrigo Ibata}
\affiliation{Universit\'e de Strasbourg, CNRS, Observatoire astronomique de Strasbourg, UMR 7550, F-67000 Strasbourg, France}
\email[show]{rodrigo.ibata@astro.unistra.fr}

\author[0000-0001-8392-3836]{Wassim Tenachi}
\affiliation{Mila - Quebec Artificial Intelligence Institute}
\affiliation{D\'epartement de physique, Universit\'e de Montr\'eal}
\affiliation{Ciela - Montreal Institute for Astrophysical Data Analysis and Machine Learning}
\email{wassim.tenachi@umontreal.ca}

\author[0000-0002-8788-8174]{Foivos Diakogiannis}
\affiliation{Technology, Commonwealth Scientific and Industrial Research Organisation (CSIRO), Kensington, WA 6151, Australia}
\email{Foivos.Diakogiannis@data61.csiro.au}

\author[0000-0003-1559-1053]{Neil Ibata}
\affiliation{Department of Human Evolutionary Biology, Harvard University, Cambridge, MA, USA}
\email{neilibata@fas.harvard.edu}

\author[0009-0008-7455-1880]{Anirudh Shankar}
\affiliation{Universit\'e de Strasbourg, CNRS, Observatoire astronomique de Strasbourg, UMR 7550, F-67000 Strasbourg, France}
\email{anirudh.shankar@astro.unistra.fr}

\begin{abstract}
Scientific analyses increasingly use flexible neural networks, but their
thousands of correlated parameters make it challenging to interpret the
associated uncertainties.  Here we develop a low-dimensional posterior for
the fitted function itself, for scientific neural surrogates trained with
second-order optimization.  Linearizing the fitting procedure with respect
to the randomized residual rows gives a measurement-to-function transport,
the linear map, assembled from the converged Jacobians and Gauss--Newton
curvature, that carries measurement perturbations into the function
perturbations that refitting would produce.  Its leading
singular functions define coherent deformation modes.  Independent Gaussian
coefficients then generate smooth function draws, so that any derived
quantity, including those requiring derivatives or integrals of the draw,
inherits the posterior.  The construction distinguishes repeated-experiment
covariance from the local Gauss--Newton/Laplace posterior and propagates
both to correlated quantities of scientific interest.  The result is conditional on
the fit's declared choices (architecture, hyperparameters, active set, and
optimization branch), and every fit is certified as converged by checking
that a further optimization step would change the fitted predictions by
less than a chosen small fraction of the measurement errors.

Our primary example is an $800$-parameter phase-space
distribution function fit for a mock stellar disk.  Four uncertainty
coordinates, two orders of magnitude fewer than the fitted parameters and
stable under refinement of the force basis, capture 99\% of the vertical-force
posterior variance, and $4\,000$ coherent draws propagate through the force,
total-density, surface-density, and frequency calculations in $0.8$~s.  The
method provides a highly efficient route to uncertainty propagation for
derivative-dependent scientific inference.
\end{abstract}

\keywords{Astrostatistics (1882) --- Bayesian statistics (1900) --- Markov chain Monte Carlo (1889) --- Uncertainty bounds (1917) --- Neural networks (1933) --- Dark matter density (354)}

\section{Introduction}
\label{sec:intro}

When a scientific model has five parameters, its posterior is a familiar
object, such as a corner plot, a covariance, or a table of intervals.  But when the model
is a neural surrogate with thousands of parameters or more, that intuition
evaporates.  The marginal error bar of one weight in the model answers no
scientific question, and the joint covariance is a dense matrix whose
individual entries have no meaning that survives a change of network
parameterization.  In practice, one often reports only the point fit and
treats its uncertainty as a separate problem, or samples the full parameter
posterior and spends considerable computation exploring gauge directions and
weakly constrained plateaus.

Here, we take a different view, that the natural posterior object for a scientific
surrogate is not a cloud of parameters but a low-dimensional basis of
coherent function deformations.  The parameters are scaffolding.  What the
data constrain, and what science consumes, is the \emph{function}, together
with the family of functions the same experiment could plausibly have
produced.  The inference chain we construct is therefore
\begin{equation*}
\begin{split}
\text{data}\;\longrightarrow\;\text{deterministic fit}
\;&\longrightarrow\;\text{transport}\\
\;\longrightarrow\;\text{deformation basis}
\;&\longrightarrow\;\text{scientific quantities}.
\end{split}
\end{equation*}
The construction of this representation requires no parameter-space
sampling.  Random draws are needed only when propagating the reduced
posterior through nonlinear scientific quantities.  Where Markov chain Monte
Carlo (MCMC) sampling appears in this paper, it samples an independently
constructed spline posterior for the same data and serves as an external
calibration of the function-space result (\S\ref{sec:rotcurve}).

Three objects give the chain its content.  First, the
\emph{measurement-to-function transport} $\mat T_X$ (\S\ref{sec:transport}),
the linear operator that carries
measurement perturbations to the function perturbations that refitting
would produce, assembled from pieces the ``NestyNet''
second-order optimizer and segmented model presented in Paper~I
\citep{NestyNet2026a} have already computed at convergence.  Second, the
\emph{posterior deformation modes} (\S\ref{sec:modes}), the singular
functions of that transport, which compress the plausible family into a
small basis, with four elements in the $P=800$ fit of
\S\ref{sec:flagship}.  They are the dominant deformations of
the \emph{inferred} function under measurement uncertainty, filtered through
the complete inference. The low-rank truncation minimizes the expected
function-space error [Eq.~\eqref{eq:mode_truncation}].  Third, the
\emph{pushforward} (\S\ref{sec:compressions}): every derived physical
quantity inherits its posterior from the same draws, with a one-solve
linearized interval for any scalar.

Why is this program practical for NestyNet-type surrogates when it would be
cumbersome for a generic deep network?  Five properties make the
construction possible: deterministic fitting, which removes algorithmic
randomness from repeated-data experiments; analytic Jacobian--vector
products throughout both the model and the downstream science; data-informed
curvature that is often compressible at the relevant posterior scale,
together with the segment-native preconditioning linear algebra (SPLA) and
Nystr\"om machinery for large networks presented in a dedicated
theory paper~\citep{NestyNet2026Ia}; differentiable scientific outputs; and the local
reparameterization invariance of the pushed-forward covariance, which
removes parameter gauge from the function-space quantities we report.

\subsection{What is new, and what is not}

Several ingredients assembled here have antecedents.  
Reading a Gauss--Newton solve as Bayesian inference in a
locally linearized model is the linearized Laplace approximation.  It has
been classical since \citet{MacKay1992} and has been revived for neural
networks, in function space, by recent
work~\citep{Khan2019,Immer2021,Daxberger2021}.  That the posterior curvature
of an over-parameterized inverse problem is effectively low rank, and that
inference can be confined to a data-informed subspace, is a mature subject
\citep{Cui2014,Spantini2015,BuiThanh2013}.  The optimality of a truncated
singular basis goes back to~\citet{EckartYoung1936}.  And differentiating an
estimator with respect to its data is the influence function and delta
method of classical statistics, and implicit differentiation of an argmin
(the minimizer of an objective) in
its modern form~\citep{Hampel1974,Blondel2022}.  We use all of these.

A useful shorthand for the algebraic core is therefore \emph{linearized
Laplace, pushed to function space, followed by a singular value
decomposition (SVD)}.  What that shorthand omits, though, is the end-to-end
inferential construction.  We differentiate with respect to every continuous
coordinate that the declared stationarity system solves for, which includes
persistent regularization, latent input corrections, calibration parameters,
and nuisance variables, and not with respect to the network likelihood
alone.  The experiments in this paper condition on the chosen architecture
and segment count, active set, evidence-prior strengths, final robust or
iteratively reweighted least squares (IRLS) weights, discrete controller
decisions, and a smooth optimization branch\footnote{A nonlinear fit generally
admits several local solutions, and the \emph{branch} is the particular
one the optimizer has converged to, followed smoothly as the data are
perturbed, in the way one follows a chosen root of a transcendental
equation.}.  A data-derived hyperparameter
that is held fixed is therefore part of the conditioning information, not an
implicitly marginalized random variable.  We keep the repeated-experiment
transport and the prior-augmented posterior transport as two distinct
objects, and from the singular functions of each we build the function-space
basis that is Karhunen--Lo\`eve-optimal~\citep{Karhunen1947,Loeve1978} under
the chosen anchor measure.  The resulting draws are complete smooth
functions, so derivatives and integrals remain meaningful, and the same
draws propagate jointly into all named scientific deliverables.  Finally, we
validate the tangent transport against both an independently constructed
spline posterior and full nonlinear perturbed-data refits, and use a
dimensionless diagnostic, the ratio of first- to second-order
function response along each posterior direction
($Q$ in \S\ref{sec:rotcurve:gauge}), to identify directions in which that tangent
approximation is unreliable before those expensive refits are performed.
Our contribution here is thus not the transport identity itself, whose
ingredients are classical response analysis, but a compact,
coherent, and falsifiable scientific posterior construction built around it.

Section~\ref{sec:rotcurve} subjects the construction to a controlled
validation on a mock rotation curve, chosen to be small enough that an
independent spline route with an exact closed-form posterior and full
perturbed-data refit ensembles are both available as references, and uses
the same setting to probe where the tangent approximation reaches its
limits.  Section~\ref{sec:flagship} scales the construction to an
$800$-parameter phase-space fit and the coherent vertical-force and
density tower built on it, and \S\ref{sec:rom} closes by asking what kind
of object the mean model and its deformation modes form together.

\section{The Measurement-to-Function Transport}
\label{sec:transport}

Every astronomer propagates errors: differentiate a formula with respect
to its measured inputs and carry the measurement covariance through the
derivative.  The object at the center of this paper is that same
construction applied to an entire fitting procedure.  A least-squares fit
is itself a map, from the measured data vector to the fitted function.
Near a converged fit that map is smooth, and its derivative is a linear
operator that carries (``transports'') a small perturbation of the
measurements to the coherent perturbation of the fitted function that
refitting would have produced.  We call this operator the
measurement-to-function transport.  To first order it answers, at
negligible cost, the question one could otherwise address only by
brute-force repeated refits: had the data scattered differently within
their quoted uncertainties, how would the inferred function have changed?

\subsection{Linear response of the fitted function}

Let $\vartheta$ collect the continuous coordinates included in the declared
fit: the network parameters and, when present, latent input corrections,
calibration parameters, or nuisance parameters.  Discrete model choices and
conditioned hyperparameters are not silently included in $\vartheta$.  We
write the objective as
\begin{equation}
\Phi(\vartheta;\tilde{\vect y})
=\tfrac12\bigl\|\tilde{\vect y}-\vect g(\vartheta)\bigr\|^2
+\tfrac12\bigl\|\vect r_c(\vartheta)\bigr\|^2,
\label{eq:objective}
\end{equation}
where $\tilde{\vect y}=\mat\Sigma_y^{-1/2}\vect y$ are the whitened
measurements ($\mat\Sigma_y$ is the measurement covariance), $\vect g$
contains the correspondingly standardized model
predictions at the measured inputs, and $\vect r_c$ collects persistent
prior, physics, and finite-penalty constraint residuals.  We denote their
Jacobians by $\mat J_m=\partial\vect g/\partial\vartheta$ and $\mat
J_c=\partial\vect r_c/\partial\vartheta$.  Unless stated otherwise, all
Jacobians below are evaluated at the converged solution $\hat\vartheta$.

The fitted parameters satisfy
$\nabla_\vartheta\Phi(\hat\vartheta;\tilde{\vect y})=\vect 0$.  Define the
exact stationary-point Hessian $\mat H_{\rm stat}=\nabla_\vartheta^2\Phi$.
If the whitened measurements change by $\delta\tilde{\vect y}$, linearizing
this stationarity condition gives
\begin{equation}
\mat H_{\rm stat}\,\delta\hat\vartheta
=\mat J_m^\top\delta\tilde{\vect y} \, .
\label{eq:fit_map_linearization}
\end{equation}
The corresponding change in the fitted function on any query set $X$ is
therefore
\begin{equation}
\delta\hat f_X
=\mat J_{f,X}\mat H_{\rm stat}^{-1}
 \mat J_m^\top\delta\tilde{\vect y} \, ,
\label{eq:exact_transport}
\end{equation}
where $\mat J_{f,X}=\partial f_X/\partial\vartheta$.
Equation~\eqref{eq:exact_transport} is the first-order response of the declared
stationary branch, which should not be conflated with the response 
of the network with its parameters held fixed.

For computation we replace $\mat H_{\rm stat}$ by the Gauss--Newton
curvature
\begin{equation}
\mat H\equiv\mat H_{\rm GN}
=\mat J_m^\top\mat J_m+\mat J_c^\top\mat J_c \, ,
\label{eq:H_def}
\end{equation}
consistent with both the optimizer and the local Laplace approximation.
The measurement-to-function transport is then
\begin{equation}
\mat T_X
:=\mat J_{f,X}\mat H_{\rm GN}^{+}\mat J_m^\top \, ,
\qquad
\delta\hat f_X\simeq\mat T_X\,\delta\tilde{\vect y} \, .
\label{eq:transport}
\end{equation}

For compactness below, $\mat H^{-1}$ denotes the ordinary inverse when the
declared curvature is full rank and the retained-subspace pseudoinverse
$\mat H^+$ otherwise.  In the singular case $\mat H^{+}$ defines a
Gaussian on the retained quotient subspace, not a proper flat-prior
posterior over all parameters.  We diagonalize the symmetrized curvature in the Euclidean
metric of the optimizer-facing parameter coordinates and retain precisely
the directions with
\begin{equation}
\lambda_k/\lambda_{\max}>\tau_{\rm eig} \, ,
\label{eq:rank_rule}
\end{equation}
where the $\lambda_k$ are the eigenvalues of that curvature,
$\lambda_{\max}$ is the largest, and $\tau_{\rm eig}$ is a dimensionless
truncation threshold.  We use $\tau_{\rm eig}=10^{-10}$ for the rotation
fits and $10^{-12}$ for the regularized flagship fit
(\S\ref{sec:flagship}).

The curvature in Eq.~\eqref{eq:H_def} is rebuilt from the converged Jacobian
and the persistent residual rows.  In particular, transient
Levenberg--Marquardt (LM) damping is excluded. It belongs to the route by
which the optimum was reached, not to the fitted statistical model.  An
evidence-controller Gaussian prior (\citealt{MacKay1992}, and adopted in Paper~I) enters
through $\vect r_c$ and therefore remains in $\mat H_{\rm GN}$.  In the
evidence experiments below the segment prior strengths are fixed at their
declared component values in every mock.  The evidence weighs competing
fixed-strength components against one another
(\S\ref{sec:flagship:evidence}) but never re-adapts a strength to an
individual mock realization, so both the transport and refits are
conditional on those empirical-Bayes choices.  Re-optimizing them per
realization would add an outer-loop response absent from
Eqs.~\eqref{eq:exact_transport}--\eqref{eq:transport}.

\subsection{Repeated experiments and the local posterior}
\label{sec:transport:flavors}

The transport first has a purely sampling-based interpretation.  Consider a
repetition of the experiment in which the whitened measurement perturbation
is $\vect\epsilon\sim\mathcal N(\vect 0,\mat I)$.  To first order,
$\delta\hat f_X=\mat T_X\vect\epsilon$, and hence
\begin{equation}
\begin{split}
\mat C_{\rm rep}(X,X')
&=\mat T_X\mat T_{X'}^\top\\
&=\mat J_{f,X}\mat H^{-1}\mat H_L\mat H^{-1}
  \mat J_{f,X'}^\top \,,\\
\mat H_L&=\mat J_m^\top\mat J_m \, .
\end{split}
\label{eq:boot_cov}
\end{equation}
This is the first-order sampling covariance of the fitted function under the
quoted measurement errors.  It is the delta method (first-order error
propagation) applied to the entire
inference pipeline and requires no Bayesian interpretation.

A different object results when the persistent regularization is interpreted
as a proper Gaussian prior with precision $\mat P$.  For clarity, first
consider the common case $\mat H=\mat H_L+\mat P$. Additional Gaussian
likelihood blocks are included by appending their whitened Jacobian factors
in the same way.  Suppressing the query label for compactness, the
repeated-experiment and posterior transports, the two \emph{flavors} of
the construction, are
\begin{equation}
\mat T_{\rm rep}
=\mat J_f\mat H^{-1}\mat J_m^\top,
\qquad
\mat T_{\rm post}
=\mat J_f\mat H^{-1}
\begin{bmatrix}
\mat J_m^\top & \mat P^{1/2}
\end{bmatrix}.
\label{eq:two_transports}
\end{equation}
The second expression gives the covariance of the local
Gauss--Newton/Laplace posterior,
\begin{equation}
\begin{split}
\mat C_{\rm post}(X,X')
&=\mat T_{{\rm post},X}\mat T_{{\rm post},X'}^\top\\
&=\mat J_{f,X}\mat H^{-1}\mat J_{f,X'}^\top.
\end{split}
\label{eq:post_cov}
\end{equation}
Equivalently, a posterior perturbation is generated by drawing
$\vect\epsilon$ and $\vect\eta$ independently from
$\mathcal N(\vect 0,\mat I)$ and solving
\begin{equation}
\mat H\vect h
=\mat J_m^\top\vect\epsilon+\mat P^{1/2}\vect\eta,
\qquad
\delta f_X=\mat J_{f,X}\vect h.
\label{eq:posterior_draw}
\end{equation}
The covariance of $\vect h$ is then $\mat H^{-1}$ in this proper
likelihood-plus-prior case.

More generally, some of the persistent constraint rows shape where the fit
lands but are not themselves random draws, as they carry no sampling noise.  If
$\mat H_{\rm rand}$ denotes the sum of the curvature blocks associated with
the rows that are actually randomized, the draw covariance is
\begin{equation}
\mat C_{\rm draw}(X,X')
=\mat J_{f,X}\mat H^{-1}\mat H_{\rm rand}\mat H^{-1}
 \mat J_{f,X'}^\top.
\label{eq:randomized_cov}
\end{equation}
Equation~\eqref{eq:boot_cov} has $\mat H_{\rm rand}=\mat H_L$, while
Eq.~\eqref{eq:post_cov} follows when all persistent curvature is supplied by
proper likelihood or prior rows, so that $\mat H_{\rm rand}=\mat H$.  A
fixed architectural penalty that is kept in $\mat H$ but is never given a
random perturbation instead yields a local uncertainty for the fitting
procedure as a whole, rather than a Bayesian Laplace posterior.

We use three names for three different objects.  The
\emph{conditional local posterior} is the Gauss--Newton/Laplace posterior on
a fixed branch and under the conditioning choices stated above.  The
\emph{repeated-estimator law} is its first-order sampling analogue on that
same branch.  The \emph{refit distribution} is the empirical output of
rerunning the declared optimizer. It may additionally reveal basin or
controller changes.  Only the first is a Bayesian posterior.  Warm refits, which refit
perturbed data starting from the converged parameters and so stay on the
same branch, primarily test the second.  Cold starts, which rerun the
declared deterministic initializer from scratch and can therefore select
a different branch, probe the third.

The two transports answer different scientific questions.
Equation~\eqref{eq:boot_cov} describes how the estimator would vary across
repeated measurements with the prior held fixed.
Equation~\eqref{eq:post_cov} describes uncertainty in the fitted function
conditional on the observed
data and the adopted prior.  They coincide only in the unregularized,
full-rank limit.  Each therefore has its own deformation basis.  Directions
prominent in the posterior basis but weak in the repeated-experiment basis
are directions whose uncertainty is set primarily by the prior rather than
by the sampling variability of the experiment.  The flagship experiment in
\S\ref{sec:flagship:refit} measures this distinction directly.

Data-derived evidence strengths, noise scales, prior anchors, scale
templates, and robust weights require an additional declaration. They must
be conditioned on, included in the joint stationarity system, or
re-estimated in every repeated experiment.  The experiments here use the
first option for evidence strengths and contain no data-updated robust
weights.  In the flagship fit below, the prior mean and
the family root-mean-square (RMS) scale template are fixed before any data
are seen.  The controller is run in its seed-anchor mode, so the anchor is the
deterministic seed initialization and the template is the family RMS of that
same anchor, both functions of the declared network topology and the fixed
random seed alone.  (Paper~I's data-dependent canonical anchor is not used
here, since it would have to be declared as conditioning information.)  
Fresh refits reconstruct the same anchors and hold the prior strengths 
$\alpha$ fixed, so no hidden prior ingredient responds to the mock
data.

Only residual rows representing stochastic information receive random
perturbations.  Measurement rows are randomized in both constructions,
proper-prior rows only in posterior draws, and deterministic penalties in
neither.  An exact equality constraint would receive no noise either and
would do more, confining every draw to its null space. No fit here imposes
one, the sole equality anywhere being the $K_z(0)=0$ ablation of
\S\ref{sec:flagship:setup}, which acts on the downstream force solve rather
than the network fit.

Applying either transport requires no explicit covariance matrix.  A draw
consists of one residual vector--Jacobian product, one curvature solve, and
one prediction Jacobian--vector product, all operations already supplied by
the optimizer.  
Finally, a smooth invertible change of coordinates on $\vartheta$
(transforming the prior and any persistent constraints to match) leaves the
full-rank pushed-forward covariance $\mat J_f\mat H^{-1}\mat J_f^\top$
unchanged, but a fixed numerical cutoff in the pseudoinverse, or a screen by
the tangent-adequacy ratio $Q$ (\S\ref{sec:rotcurve:gauge}) applied in
parameter directions, can pick out a different subspace once the coordinates
are rescaled.  We therefore report the sensitivity to that cutoff and judge
convergence by its effects in function space.  This is why the uncertainties
reported below live in function and deliverable space rather than in
individual network coordinates.

\section{Posterior Deformation Modes}
\label{sec:modes}

The transport maps a standard-normal perturbation in measurement or
posterior-noise space to a coherent perturbation of the fitted function.  On
anchor points $X$ with positive quadrature/mass matrix $\mat W_\mu$ (the
discrete weights defining the $L^2(\mu)$ inner product on $X$, with $\mu$
the declared anchor measure), define
\begin{equation}
\begin{split}
\mat W_\mu^{1/2}\mat T_X &=\mat U\mat S\mat V^\top,\qquad
s_k=\mat S_{kk},\\
\psi_k(X)&=\mat W_\mu^{-1/2}\vect u_k,
\end{split}
\label{eq:weighted_svd}
\end{equation}
where $\vect u_k$ is the $k$th column of $\mat U$, the $k$th left
singular vector.  Then $\psi_i^\top\mat W_\mu\psi_j=\delta_{ij}$ for either transport
flavor (\S\ref{sec:transport:flavors}).  The experiments use uniform quadrature on
their stated one-dimensional output grids.  When several outputs are stacked
into a vector, $\mat W_\mu$ is block diagonal, one block per output, after
each output has been divided by its stated characteristic scale.  The anchor
measure and the output metric are thus declared parts of the representation,
not hidden defaults.  For the corresponding standard-normal input
perturbation $\vect\xi$,
\begin{equation}
\delta\hat f
=\sum_k s_k \zeta_k\psi_k,
\qquad
\vect\zeta=\mat V^\top\vect\xi\sim\mathcal N(\vect 0,\mat I).
\label{eq:mode_expansion}
\end{equation}
The mode coefficients are therefore independent standard normals.
Truncating the expansion after $K$ terms leaves expected squared $L^2(\mu)$
error
\begin{equation}
\expect\left\|
\delta\hat f-\sum_{k=1}^{K}s_k\zeta_k\psi_k
\right\|_\mu^2
=\sum_{k>K}s_k^2.
\label{eq:mode_truncation}
\end{equation}
By the Eckart--Young--Mirsky theorem~\citep{EckartYoung1936}, equivalently
the Karhunen--Lo\`eve property of the pushed-forward covariance, no other
$K$-dimensional linear basis has a smaller expected truncation error.
Because $\mat T$ is the derivative of the composed data-to-function map,
the resulting function-space basis is unchanged by a smooth
reparameterization of the internal network coordinates, before numerical rank
truncation and with the prior and any persistent constraints transformed
consistently (the caveat of \S\ref{sec:transport:flavors} on the fixed
pseudoinverse cutoff continues to apply).

More generally, for any declared linear output operator $\mat L$ with
positive output metric $\mat W$, the weighted SVD of
$\mat W^{1/2}\mat L\mat T$ gives the Karhunen--Lo\`eve-optimal basis in that
operator-weighted metric. Derivative-weighted Sobolev norms and joint
value--derivative metrics are special cases.

Everything the transport machinery has produced now condenses into a
single formula.  The truncated expansion
\begin{equation}
\;F_{\vect\zeta}(x)=\hat f(x)+\sum_{k=1}^{K}s_k\,\zeta_k\,\psi_k(x),\qquad
\vect\zeta\sim\mathcal N(0,\mat I_K)\;
\label{eq:sister}
\end{equation}
is the \emph{generative sister model}, the mean model of Paper~I plus a
$K$-dimensional family of coherent deformations.  It need not be distilled
into a second neural network.  If $\mat B$ is the randomized-row factor
($\mat J_m^\top$ for repeated experiments, or $[\mat J_m^\top,\mat P^{1/2}]$
for posterior draws), each right singular vector defines the parameter
tangent
\begin{equation}
\vect q_k=\mat H^{+}\mat B\vect v_k,
\qquad
s_k\psi_k(x)=\mat J_f(x)\vect q_k.
\label{eq:mode_tangent}
\end{equation}
The same tangent gives analytic values and input derivatives at arbitrary
new points through the corresponding Jacobian--vector product.

It is worth being precise about what the $\psi_k$ are \emph{not}.  They are
not principal components of the data.  They are not Karhunen--Lo\`eve modes
of a prior process.  They are not eigenvectors of the parameter covariance
or of the loss Hessian, and they do not live in parameter space at all.
Instead, they are
\begin{quote}
\emph{the dominant ways the declared randomized uncertainty can deform the
inferred function on the declared stationary branch.}
\end{quote}
For the repeated-experiment flavor that randomized uncertainty is the
measurement noise alone, while for the posterior flavor it also includes the
proper-prior rows.
The filtering through inference matters on both sides.  Two large,
anticorrelated parameter excursions that cancel in the prediction produce no
mode.  A coordinated displacement of hundreds of parameters that shifts one
scientific feature becomes a single visible $\psi_k$.  And a draw of
$\vect\zeta$ is
an entire coherent curve, with derivatives and integrals of $F_{\vect\zeta}$ that are meaningful,
in sharp contrast to a heteroscedastic head\footnote{A network output layer that
predicts its own error bar at each point.} that samples $y$ independently at
each $x$, whose derivatives are nonsense and whose integrals acquire
white-noise fur.  Section~\ref{sec:rotcurve:deliv} uses exactly this
coherence when three of its five deliverables differentiate every draw.

The retained \emph{subspace} is interpretable in a way parameter covariance
never is.  In the rotation-curve experiment below it contains quasi-local outer-slope and
interior wiggle distortions, so the error budget can be read in function
space.  We do not attach fixed physical names to individual vectors when
their singular values are nearly equal, since within such a plateau any orthogonal
rotation of the vectors is equally valid, unless a further rule for choosing
the rotation is declared.

\section{Pushforward: Scientific Quantities Inherit Their Posterior}
\label{sec:compressions}

Let $M=\mathcal M(\vartheta)$ be a scalar compression of the fit (a single
physical number read off the fitted curve, such as a mass or a frequency, and in
general a nonlinear functional of $F_{\vect\zeta}$ and its derivatives) and $\vect
m=\nabla_\vartheta\mathcal M|_{\hat\vartheta}$ its gradient, available in
closed form through the same chain of Jacobian--vector products that
yields the predictions themselves.  For arbitrary randomized residual blocks
the scalar draw variance is
\begin{equation}
\operatorname{Var}_{\rm draw}(M)
=\vect m^\top\mat H^{-1}\mat H_{\rm rand}\mat H^{-1}\vect m.
\label{eq:one_solve_general}
\end{equation}
For the proper posterior case $\mat H_{\rm rand}=\mat H$,
\begin{equation}
\;\sigma_M^2=\vect m^\top\mat H^{-1}\vect m\;
\label{eq:one_solve}
\end{equation}
costs a single curvature solve regardless of the parameter count, and
$\operatorname{Cov}(M_a,M_b)=\vect m_a^\top\mat H^{-1}\vect m_b$ gives the
fully correlated posterior of several deliverables at once.

The repeated-estimator analogue is
\begin{equation}
\operatorname{Var}_{\rm rep}(M)
=\vect m^\top\mat H^{-1}\mat H_L\mat H^{-1}\vect m
=\|\mat J_m\mat H^{-1}\vect m\|^2,
\label{eq:one_solve_repeated}
\end{equation}
which likewise costs one solve followed by a data-Jacobian product.  In
either flavor the uncertainty of one scalar occupies a single coordinated
direction $\mat H^{-1}\vect m$ through the parameter labyrinth, so the
hopeless-looking correlations combine automatically.  The gradient $\vect m$ may involve
input derivatives of the fit (the epicyclic frequency of the
rotation-curve benchmark in \S\ref{sec:rotcurve}, for example, needs
$\partial_r F_{\vect\zeta}$),
and the mixed derivative with respect to both radius and the network
parameters is available in closed form, just like the function itself.

Beyond the linearization, the sister simply \emph{is} the pushforward
sampler: $\vect\zeta_s\sim\mathcal N(0,\mat I_K)$, $M_s=\mathcal M[F_{\vect\zeta_s}]$, and the
quantiles of $\{M_s\}$ are the credible interval.  The compression's own
nonlinearity (a squared velocity, a root, an integral bound) is retained
exactly.  Only the $\vartheta\to f$ map is linearized, an approximation
\S\ref{sec:rotcurve:refit} tests directly against refits.

A pointwise credible envelope for the underlying function comes from
$F_{\vect\zeta}$ alone.  A
prediction interval for a future noisy datum adds fresh detector noise in a
separate observation layer,
$\operatorname{Cov}(Y_i,Y_j\mid D)=\mat
C_f(x_i,x_j)+\delta_{ij}\sigma_{y,i}^2$, with $\mat C_f$ the function
covariance of the chosen flavor, $D$ the observed data, and
$\sigma_{y,i}^2$ the $i$th measurement variance.  Correlated detector
noise replaces the diagonal term by $(\mat\Sigma_y)_{ij}$.

\section{A Controlled Validation Suite on a Rotation Curve}
\label{sec:rotcurve}

\begin{figure*}[t]
\centering
\includegraphics[width=\textwidth]{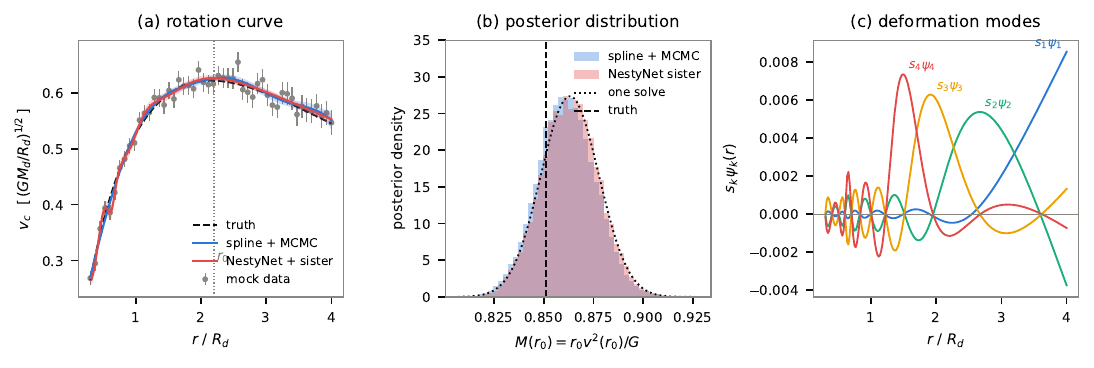}\\[8pt]
\includegraphics[width=0.92\textwidth]{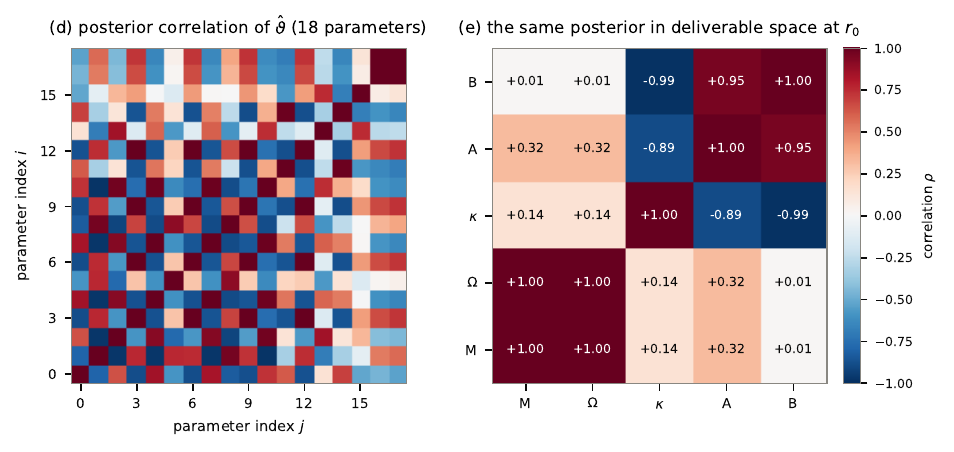}
\caption{Controlled validation of the sister on a mock rotation curve, and
  one local posterior represented in two languages.
  (a)~$N=50$ mock velocities from a razor-thin exponential-disk rotation
  curve~\citep{Freeman1970} (dashed), with heteroscedastic measurement
  errors, and the pointwise 68\% bands on the underlying (latent) function
  obtained by two independent methods: a cubic B-spline sampled by ensemble
  MCMC (blue) and a NestyNet fit with the Laplace sister (red).  The dotted
  line marks $r_0=2.2\,R_d$.  (b)~The pushforward posterior of the
  spherical-equivalent dynamical-mass proxy $M(r_0)=r_0v^2(r_0)/G$: MCMC
  histogram (blue), sister draws (red), the one-curvature-solve Gaussian of
  Eq.~\eqref{eq:one_solve} (dotted), and the analytic truth (dashed).
  (c)~One valid orthonormal choice of the leading block of
  near-degenerate deformation modes, plotted as $s_k\psi_k(r)$
  [Eq.~\eqref{eq:sister}]. The singular values
  $(s_1,\ldots,s_4)=(0.041,0.038,0.034,0.032)$ are close, so only their span
  is invariant, not the plotted vectors or their physical names.
  (d)~Posterior correlation matrix of the same fit's 18 parameters, dense
  (median $|\rho|=0.65$, with 13\% of pairs above $0.9$) and with no entry
  that survives a change of network parameterization.  (e)~The identical
  posterior pushed to deliverable space at $r_0$ ($M$, $\Omega$, $\kappa$,
  $A$, $B$), from one set of
  $20\,000$ coherent draws ($\kappa$, $A$, $B$ differentiate every draw).
  The structure reads as physics, e.g.\ $\rho(\kappa,B)=-0.99$.}
\label{fig:rotcurve}
\end{figure*}

The rotation curve below is deliberately \emph{not} a hard problem.  It is
chosen as a calibration chamber, small enough that an independent, fully
conventional route to the function-space posterior exists (a spline model,
linear in its coefficients, whose posterior is exactly Gaussian in closed
form, sampled both
exactly and by an affine-invariant ensemble
MCMC~\citep{GoodmanWeare2010,ForemanMackey2013}) and small enough that full
perturbed-data refit ensembles are cheap.  The sister is required to agree
with both.

\subsection{Setup}

Mock data are drawn from the circular velocity of a razor-thin
exponential disk~\citep{Freeman1970,BinneyTremaine2008},
$v_c^2(r)=4\pi G\Sigma_0R_d\,y^2[I_0(y)K_0(y)-I_1(y)K_1(y)]$,
with dimensionless radius $y=r/2R_d$, central surface density $\Sigma_0$, 
exponential scale length $R_d$, disk mass $M_d$, and modified 
Bessel functions $I_n$ and $K_n$. We use units $G=M_d=R_d=1$, 
and take $N=50$ radii evenly spaced on $[0.3,4]\,R_d$ with Gaussian errors
$\sigma_i=0.012+0.010\,(r_i/4R_d)$.  The conventional route fits a cubic
B-spline (8 basis functions, flat prior).  The NestyNet route
fits a six-segment scalar single-layer model with $P=18$ parameters
(weighted $\chi^2/\mathrm{dof}=18.1/32$) and
builds the Laplace sister (in 1\,ms) from the converged Jacobian
[Eqs.~\eqref{eq:H_def}, \eqref{eq:post_cov}].  Both
routes push the same spherical-equivalent dynamical-mass proxy,
$M_{\rm sph}(r_0)=r_0v^2(r_0)/G$ (hereafter $M$), at
$r_0=2.2\,R_d$ (true value $0.8512$), defined as a functional of the
curve.

\subsection{Agreement with the independent route, and calibration}

In Figure~\ref{fig:rotcurve}b, two different model
classes and two different inference engines return the same proxy posterior,
to a few percent in width and well within a standard error in center, and
both cover the truth.  For the plotted realization (seed 42, truth $0.8512$)
the two spline routes give $M(r_0)=0.8612$ with 68\% interval
$[0.8471,0.8758]$ under ensemble MCMC and $0.8618$ with $[0.8470,0.8766]$
from the exact closed-form posterior, while the sister gives $0.8631$ with
$[0.8488,0.8774]$ from draws and $0.8631\pm0.0146$ from the one solve of
Eq.~\eqref{eq:one_solve}. The four interval widths span
$0.0286$--$0.0296$.  Over 20 independent mock realizations the 68\%
intervals cover the truth in 16 of 20 (spline) and 14 of 20 (sister)
realizations, both consistent
with nominal coverage at the binomial resolution ($\pm0.10$) of the test.

This 20-realization experiment is a calibration smoke test (a coarse
first-pass sanity check).  After the fit, the
sister's draws cost $0.14$~s and the one-solve interval $0.4$~ms.  The toy
MCMC is of course also cheap.

\subsection{The deliverable-space posterior}
\label{sec:rotcurve:deliv}

What does this construction say that parameter covariance
cannot?  Figures~\ref{fig:rotcurve}d and \ref{fig:rotcurve}e give the
concrete answer.  From the \emph{same} fit and the \emph{same} $20\,000$
draws, the sister simultaneously delivers the posterior of five physically useful
quantities at $r_0$: the mass estimator $M=r_0v^2/G$, the angular frequency
$\Omega=v/r_0$, the epicyclic frequency $\kappa=[2\Omega(\Omega+v')]^{1/2}$,
and the Oort constants $A=\tfrac12(\Omega-v')$, $B=-\tfrac12(\Omega+v')$
\citep{BinneyTremaine2008}.  Three of the five require the radial derivative
of every draw. This is where the coherence of Eq.~\eqref{eq:sister} stops
being aesthetics and starts being load bearing, and where the segmented
models' analytic derivative with respect to both radius and the network
parameters becomes profitable.  In this realization each of the five marginal
intervals contains its truth ($M$: $0.863\pm0.014$ vs $0.851$; $\Omega$:
$0.2847\pm0.0024$ vs $0.2827$; $\kappa$: $0.401\pm0.011$ vs $0.397$; $A$:
$0.143\pm0.008$ vs $0.143$; $B$: $-0.141\pm0.007$ vs $-0.139$), and the
correlation matrix is legible on sight, showing a $v(r_0)$ block and a $v'(r_0)$
block, with the correlation $\rho(M,\Omega)$ rounding to unity for the narrow
positive draws (the nonlinear relation is not exactly affine).  The
parameter correlation matrix of the identical posterior
(Figure~\ref{fig:rotcurve}d)
communicates none of this.  Parameter covariance contains the information,
but it is the space of deliverables that can render it intelligible.

\subsection{The near-gauge boundary, and a diagnostic that flags it}
\label{sec:rotcurve:gauge}

Our first implementation of the pushforward re-evaluated the full network at
each perturbed parameter vector, and it failed spectacularly: the $M$
interval inflated to $[0.47,\,8.50]$.  The diagnosis is instructive.  The
fitted curvature spectrum spans ten decades
($\lambda\in[1.6\times10^{-1},\,6.7\times10^{8}]$). Its lowest retained
directions are nearly unconstrained redundancies among the segments of the
one-dimensional segmented model (directions that are almost pure gauge), with
$1\sigma$ draw amplitudes of order unity.  Along such a direction the
response is purely quadratic: a $\pm1\sigma$ excursion moves $f(r_0)$
from $0.626$ to $1.42$ and $1.36$, both up.

This boundary deserves clarification.  Function-space linearization
removes parameter gauges exactly, but it distinguishes a true gauge from a
weakly identified direction only to first order.  Quadratic response along
weakly identified directions is a non-Gaussian effect, and must be assessed
by refits, higher-order transport, or an explicit prior.  The linearized
sister is the correct local tangent posterior over functions. It does not
manufacture the nonlinear response of unresolved directions, and no
covariance trick can convert an unconstrained excursion into a measurement.

The useful news is that the boundary is \emph{detectable in advance}.  For
each retained curvature direction $\vect v$ with posterior scale
$\sigma_v=\lambda_v^{-1/2}$, define the dimensionless tangent-adequacy ratio
\begin{equation}
Q(\vect v)=
\frac{\sigma_v\,\|\mat J_f\vect v\|_\mu}
{\tfrac12\,\|f(\hat\vartheta+\sigma_v\vect v)
+f(\hat\vartheta-\sigma_v\vect v)-2\hat f\|_\mu},
\label{eq:adequacy}
\end{equation}
the first- over second-order function response at one posterior standard
deviation (two extra model evaluations per direction).  On the fiducial fit
the linear response is nearly constant across all twelve retained directions
($\sigma_v\|\mat J_f\vect v\|_\mu\approx 0.002$) while the quadratic
response spans five decades. The ratio $Q$ rises from $2\times10^{-3}$ to
$5\times10^{8}$, and the five directions flagged as tangent-inadequate
($Q<1$, with values $0.00$--$0.33$) are precisely the directions that
produced the $[0.47,\,8.50]$ explosion.  We evaluate $Q$ on retained
Euclidean curvature eigenvectors, with an absolute effect floor for
numerically null responses.  The test is therefore basis-dependent and
diagnostic. A direction with $Q<1$ is flagged for a prior, cutoff audit, or
refit, but never silently removed.  We report the unscreened failure
above rather than tuning a cutoff to the known answer.  The diagnostic runs
at build time and converts an alarming failure mode into a checked
precondition.

\subsection{Certifying convergence}
\label{sec:rotcurve:cert}

The transport describes a fully converged solution.  We therefore certify the original fit and
every refit as converged by the size of its remaining Gauss--Newton
correction.  For $R$ standardized data residuals, let $\vect
h=\mat H^+\nabla_\vartheta\Phi$, with the gradient including the
persistent-prior rows.  We require
\begin{equation}
\begin{aligned}
\eta_{\rm N}&=\sqrt{(\nabla_\vartheta\Phi)^\top\mat H^+\,\nabla_\vartheta\Phi/R}\le10^{-3},\\
\eta_f&=\|\mat J_m\vect h\|/\sqrt R\le10^{-3}.
\end{aligned}
\label{eq:stationarity_certificate}
\end{equation}
$\eta_{\rm N}$ is a dimensionless retained Newton score and $\eta_f$ is the
RMS first-order change in the standardized training predictions.  A fit that fails the
certificate is discarded rather than admitted to a refit ensemble
(\S\ref{sec:rotcurve:refit}).  We separately check that the gradient along
the weakly constrained directions dropped by the eigenvalue cutoff is
negligible, rejecting a fit when it exceeds
$10^{-3}\,\|\nabla_\vartheta\Phi\|$ above a
roundoff floor, and we report the sensitivity to that cutoff because it is a
relative threshold that shifts under reparameterization.  Lowering the
fiducial rotation-curve cutoff by a decade leaves the $\kappa$ width
unchanged, while raising it by a decade reduces that width by 22\%.

\subsection{Perturbed-refit validation of the transport}
\label{sec:rotcurve:refit}

\begin{figure*}[t]
\centering
\includegraphics[width=\textwidth]{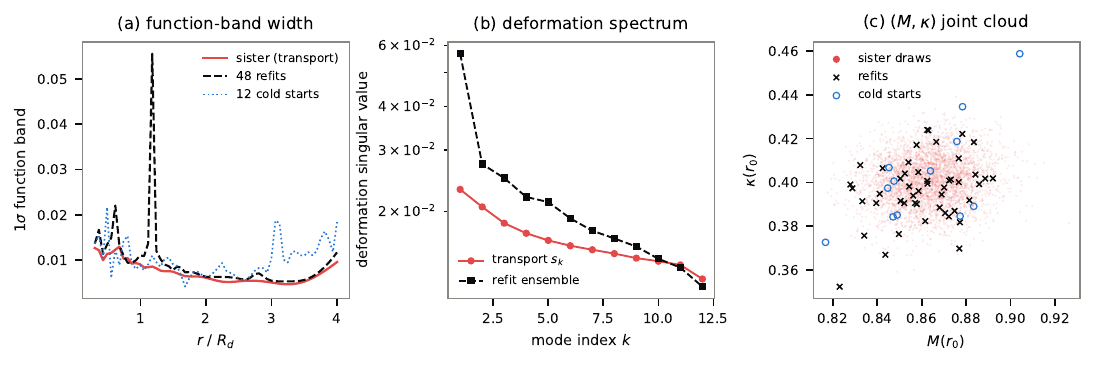}
\caption{The transport tested against fully converged nonlinear refits
  (\S\ref{sec:rotcurve:refit}): 48 parametric mock datasets refit warm from
  the original best-fit parameters, so that they stay on the same solution
  branch, with a paired subset of 12 also refit from the standard fixed
  starting point.  Every fit must pass
  Eq.~\eqref{eq:stationarity_certificate}.  (a)~Function-band standard
  deviations, refits versus transport.  (b)~Deformation singular values, warm
  refits versus transport.  (c)~The joint $(M,\kappa)$ cloud: transport draws
  (red), warm refits (crosses), and paired cold starts (blue circles).}
\label{fig:refit}
\end{figure*}

Throughout this paper a ``95\% finite-refit interval'' is a nonparametric
percentile interval from $10^4$ bootstrap resamples.  The resampling unit is
one coherent refit profile. Warm and cold ensembles are resampled
separately.  Monte Carlo error in the much larger sister reference is not
included.  The same convention is used for the flagship refits.

The central approximation of the sister is not that scientific compressions
are linear (their nonlinearity is retained), but that the stationary fitting
map $\tilde{\vect y}\mapsto\hat f$ is locally tangent, so that refitting
perturbed data and applying $\mat T$ agree to first order over perturbations
the size of the quoted noise.  We test this with 48
mocks $y_b=\hat f(x)+\gamma\sigma\epsilon_b$, with $\epsilon_b$ a standard
normal draw and $\gamma$ a noise-rescaling factor ($\gamma=1$ is the
quoted noise), each fit on a fresh,
isolated model instance (Figure~\ref{fig:refit}).
Warm starts at $\hat\vartheta$ test the same local branch.  Twelve paired
cold starts rerun the fixed-seed random initializer on each mock.  
All other topology and optimizer settings are held fixed.
The cold ensemble therefore probes only the deterministic initializer's
branch selection under the same declared estimator.

At $\gamma=1$, the warm function-band ratio is $1.132$ with 95\% interval
$[1.017,1.217]$, while the cold ratio is $1.296$ with the wider finite-ensemble
interval $[1.019,1.447]$.  The warm deformation ensemble leaves $13.9\%$ of
its centered variance outside the rank-12 transport subspace; equivalently,
the variance-weighted projection overlap is $86.1\%$.

Thus the local transport is useful on a fixed branch, while the full fitting
procedure carries additional branch and path scatter.

The derivative deliverable sharpens the same verdict.  For $\kappa$ the
transport prediction is stable at $0.0107$ under every internal variation
we tried (direct delta, full resolved transport, value modes retaining
95\% or 99\% of the variance), whereas warm refits give $0.0145$ (ratio
$1.356$, 95\% interval $[1.021,1.665]$) and cold refits $0.0243$ (ratio
$2.274$, interval $[1.036,3.010]$).  Since every variant agrees, neither
mode truncation nor a reweighted metric causes the deficit.  Neither does
the Gauss--Newton approximation.  Rebuilding the transport from the exact
stationary Hessian $\mat H_{\rm stat}=\mat H_{\rm GN}+\sum_i r_i\nabla^2
r_i$ leaves the median function width essentially unchanged (ratio
$0.9975$) and makes the $\kappa$ width slightly \emph{smaller}, and the
refit mocks are in any case centered on $\hat f$, where the
residual-curvature term vanishes identically.  The remaining candidates
are cutoff response, tangent nonlinearity, and optimization-branch
response.

Scaling the noise ($\gamma\in[1/8,2]$) gives residual-to-signal RMS ratios
$0.30$--$0.38$ across the range.  This non-vanishing small-noise
discrepancy, together with the cold-start result, prevents us from
identifying all of the gap with ordinary second-order response on a single
solution branch (one local optimum of the fit).  Within the declared rank the stationary warm-refit median is
reproduced at the 10--15\% function-width level. Outside it, cutoff
sensitivity, the $Q$ diagnostic, and explicit refits determine whether a
prior or a multi-branch treatment is required.  We do not try to propagate
uncertainty through the optimizer's stopping rule itself.  Instead,
Eq.~\eqref{eq:stationarity_certificate} guarantees by construction that
whatever the stopping rule leaves undone has a negligible effect on the
fitted function.

\subsection{Effective dimension}
\label{sec:rotcurve:kdim}

Define
\begin{equation}
K_\varepsilon=\min\Bigl\{K:\;
\textstyle\sum_{k\le K}s_k^2\;\ge\;(1-\varepsilon)\sum_k s_k^2
\Bigr\}.
\label{eq:keps}
\end{equation}
On the fiducial fit, $K_{1\%}=12$ at $P=18$.  Doubling the network to $P=36$
on the same data leaves $K_{1\%}=13$.  Thus $K_\varepsilon$ is not set by
parameter count alone, though compression is modest at this toy scale
because fifty data points give the modes nearly equal weight.  The 
boundary appears to be at $S=24$ segments ($P=72>N$): the unregularized fit interpolates
($\chi^2=3.5$) and $K_{1\%}$ grows to $28$.
Without a prior that limits model complexity, such a ``posterior'' reflects
the freedom of the model rather than anything the
data measure.  The evidence controller of Paper~I supplies
exactly that prior, as the flagship of \S\ref{sec:flagship} will
demonstrate at scale.  Thus
$K_\varepsilon$ depends jointly on the data, prior, model class, anchor
metric, and selected deliverable.  The deliverables make this concrete: all
five local rotation-curve quantities depend only on the pair
$(v(r_0),v'(r_0))$ (a two-component jet), so their coherent tangent
posterior can be generated
exactly from at most two Gaussian coordinates, even though the global
value-space curve posterior needs twelve modes to contain $99\%$ of its
variance.

The controlled tests of this section establish local transport accuracy
and expose where it fails. They do not turn what is a Laplace
approximation, conditional on the declared model, built from a single fit
branch, and linearized, into a complete accounting of every source of
uncertainty.  The rotation-curve set of coverage tests contains only 20
realizations and is therefore a smoke test limited by that small sample
size, while the refit width ratios come with explicit finite-ensemble
error intervals.  The construction does not marginalize over unlisted
architectures, conditioned hyperparameters, robust-weight updates, or
disconnected basins.

\section{The Flagship: A Vertical-Force and Local-Density Tower}
\label{sec:flagship}

While Figure~\ref{fig:rotcurve} showed the power of the method to
provide credible intervals on physically interesting quantities,
a more memorable case is the regime where a full parameter-space
representation is scientifically inefficient.  To demonstrate this we
choose as our example a $\sim\!10^3$-parameter fit of a surface
representing the phase-space distribution function (the density of stars
in position and velocity) of a galactic disk and, from that one fit,
deliver the entire coherent tower of vertical force, total dynamical
density, surface density, and vertical frequency, in under a second.  

\subsection{Setup: a thousand-parameter phase-space surface}
\label{sec:flagship:setup}

We reuse Paper~I's vertical-force demonstration (\S9), a mock
self-gravitating isothermal slab of scale height $a$ and velocity dispersion
$\sigma$, with potential $\Phi(z)=2\sigma^2\ln\cosh(z/a)$ at height $z$ above
the plane, from which $5\times10^5$ stars are drawn from the stationary
phase-space distribution function $f\propto e^{-(v^2/2+\Phi)/\sigma^2}$ in
vertical velocity $v$ (units $G=\sigma=a=1$).  We fit $\hat g=\ln\hat f$ as a segmented surface over the
phase plane $(z,v)$ to the binned counts through their exact Poisson deviance
likelihood, at a high capacity of $100$ segments.  With one output, two inputs, and
Paper~I's segment width $W=2$, each segment carries $W$ soft hinges of four
parameters (two kernel weights, a bias, an amplitude), so $P=800$ (with the
SPLA theory paper's atomic width-one counting this corresponds to $200$ hinges).  The
collisionless Boltzmann equation (CBE) then gives a chain of deliverables
that depend linearly on a smooth force field $K_z(z)=\sum_\ell\beta_\ell
B_\ell(z)$ read off the fitted surface.  This force field (not the fit) is
expanded in the $B_\ell$ using eight cubic B-splines with clamped endpoint knots on
$[-a,a]$ and no additional boundary equality.  The fiducial calculation
deliberately does not impose the known plane-symmetry condition $K_z(0)=0$,
so that it tests the derivative surrogate rather than receiving the mock
answer as a boundary condition.

At collocation points $(z_k,v_k)$ (the phase-space samples at which the
CBE rows are evaluated), the CBE and its $z$ derivative give the
two linear blocks (subscripts on $\hat g$ denoting its partial
derivatives)
\begin{equation}
\begin{split}
(\mat A_0)_{k\ell}&=\hat g_v(z_k,v_k)B_\ell(z_k),\qquad
 (\vect b_0)_k=-v_k \hat g_z(z_k,v_k),\\
(\mat A_1)_{k\ell}&=\hat g_v B'_\ell+\hat g_{zv}B_\ell,\qquad
 (\vect b_1)_k=-v_k \hat g_{zz}.
\end{split}
\label{eq:cbe_gls}
\end{equation}
We solve the stacked system with equal block weights ($w_1=1$) and a
$10^{-8}$ Tikhonov ridge scaled by the mean normal-matrix diagonal (the
block-weight sensitivity is audited below).

The coefficient vector $\vect\beta$ is re-solved for every sister draw, so
uncertainty in the derivative fields propagates through the nonlinear
tower solve rather than being frozen at its point estimate.
\begin{equation}
\begin{gathered}
\hat g(z,v)\;\to\;\{\hat g_z,\hat g_v,\hat g_{zz},\hat g_{zv}\}
\quad{}\\[-2pt]
\xrightarrow{\text{CBE least squares}}\;
K_z\to K_z'\to\{\rho,\Sigma,\nu\},
\end{gathered}
\label{eq:cbe_chain}
\end{equation}
with $\rho=-K_z'/4\pi G$ (Poisson),
$\Sigma_{\rm enc}(|z|)=-\operatorname{sgn}(z)K_z(z)/2\pi G$ (Gauss),
and vertical frequency $\nu^2=-K_z'=4\pi G\rho$.  The density-sensitive
derivative $K_z'$ is constrained by appending the $z$-differentiated CBE, so
it uses only the analytic second derivatives of $\hat g$ (Paper~I, \S9).
Profiles are tabulated on an evenly spaced 60-point grid over $[-a,a]$, while all
headline midplane scalars are evaluated continuously and exactly at $z=0$.

At this capacity the point fit is a cautionary tale on its own.  The
unregularized fit reaches the per-bin noise floor
($\mathrm{deviance}/\mathrm{dof}\approx0.95$) yet never passes the
stationarity certificate of Eq.~\eqref{eq:stationarity_certificate}, even
with ten times the optimization budget.  Its positive-semidefinite
likelihood curvature $\mat H_L$ is numerically rank deficient (rank
$96/800$ at $\tau_{\rm eig}=10^{-12}$, a count that shifts with both the
eigenvalue cutoff and the optimization budget), and the unresolved null
space makes the unregularized flat-prior Laplace posterior improper. It
does not normalize, and no finite covariance exists over the full
parameter set.  The point estimate is also unstable in the manner of the
interpolation law of Paper~I. Its $K_z$ root-mean-square error is
$0.014\,\sigma^2/a$ at the adopted optimization budget but grows to
$0.057\,\sigma^2/a$, four and a half times the regularized value, when the
optimizer is run ten times longer into the interpolation regime, even as
the deviance keeps falling.  Paper~I's evidence-based regularization
repairs all of this at once.  Fitting
with the evidence prior recovers $K_z$ to a root-mean-square error of
$0.012\,\sigma^2/a$ across the core $|z|\le a$, and the extra curvature
contributed by that persistent prior, $\mat P$,
turns the singular $\mat H_L$ into a full-rank posterior curvature $\mat
H=\mat J_m^\top\mat J_m+\mat P$ (rank $800/800$ at $\tau_{\rm
  eig}=10^{-12}$, unchanged over a decade on either side, with condition number
$8.5\times10^9$).  The prior uses one seed-anchored Gaussian block per segment (relative scale
$0.25$, floor $10^{-3}$), and its strength is set by the data through the
Laplace evidence (\S\ref{sec:flagship:evidence}), which prefers
$\alpha\simeq0.125$ independent of sample size.  Figures~\ref{fig:flagship_tower} and \ref{fig:flagship_refit} and the
four-mode result below all use that evidence-selected value.

The sister is then built from that $\mat H$, with the segment count held
fixed, no segment pruned, the evidence-selected $\alpha$ frozen, and
$\mat P$ imported directly from the evidence controller.  The persistent
prior is not a convenience here but what makes the posterior exist at
all.

Two ablations isolate choices in the CBE solve.  Changing the relative
differentiated-block weight from $w_1=1$ to $0.5$ and $2$ moves the exact
midplane mean by less than a tenth of its statistical width. Imposing the
exact equality $K_z(0)=0$ by a Karush--Kuhn--Tucker (KKT) solve drives the
maximum absolute force at zero from $1.9\times10^{-2}$ to $3.1\times10^{-16}$
over $4\,000$ draws while leaving $\rho(0)$ unchanged within its statistical
error.

\subsection{The whole tower, in under a second}
\label{sec:flagship:tower}

\begin{figure*}[t]
\centering
\includegraphics[width=\textwidth]{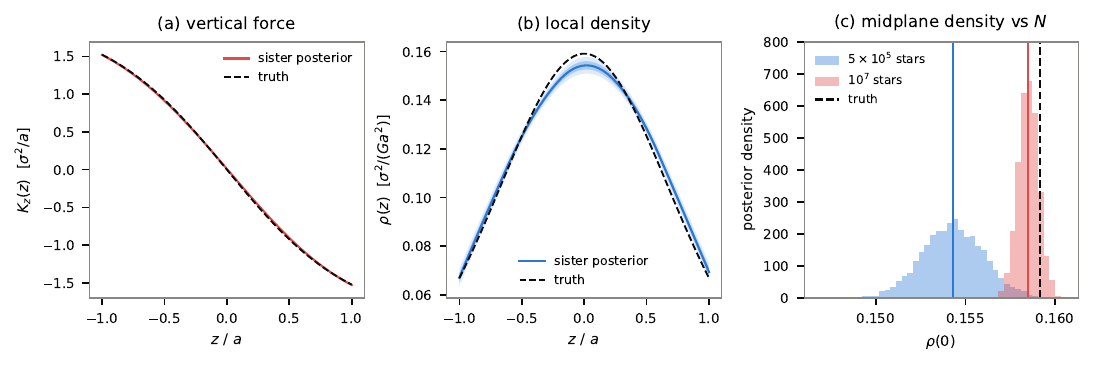}\\[8pt]
\includegraphics[width=0.92\textwidth]{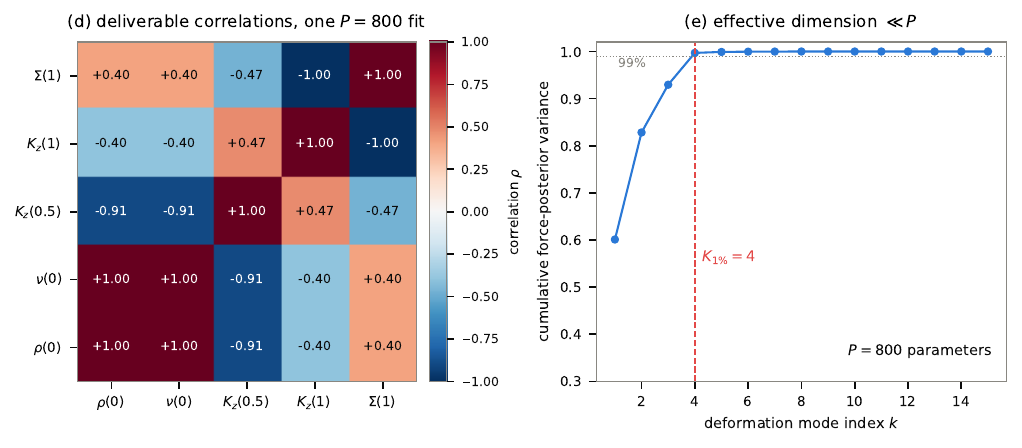}
\caption{The vertical-force and local-density tower from the evidence-selected
  $\alpha=0.125$, $P=800$ exact-Poisson fit, and the same posterior in physics
  coordinates.
  (a)~The vertical force $K_z(z)$ with the sister's $68\%$ and $95\%$
  statistical posterior bands (red) against the analytic truth (dashed).
  (b)~The local density $\rho(z)=-K_z'(z)/4\pi G$, the quantity needing the
  highest-order derivative of the fitted surface.  The statistical band is
  tight, and the small residual offset near the midplane is the
  derivative-gap systematic rather than measurement noise.  (c)~The exact-midplane
  $\rho(0)$ posterior at the survey scale ($5\times10^5$ stars) and at $10^7$
  stars against the analytic truth (dashed). The residual offset falls from
  $2.8\,\sigma$ to $1.2\,\sigma$ as the data grow, so it vanishes with sample
  size rather than being a fixed floor.  (d)~The correlation matrix of five
  named deliverables at the midplane and at the edge, from one $P=800$ fit and
  one set of draws:
  $\rho(0)$, $\nu(0)$, $K_z(0.5)$, $K_z(1)$, and $\Sigma(1)$, whereas the
  $800\times800$ parameter correlation matrix would be unintelligible.
  (e)~Cumulative variance of the $K_z$ force posterior against
  deformation-mode index. Four modes carry $99\%$.}
\label{fig:flagship_tower}
\end{figure*}

After the fit, the sister builds in $0.4$~s and pushes $4\,000$
coherent draws through the entire CBE chain of Eq.~\eqref{eq:cbe_chain} in
$0.8$~s, yielding the joint posterior of every deliverable at once
(Figure~\ref{fig:flagship_tower}).  Each draw is a coherent latent force
field, so its derivatives and integrals are physical objects. The density $\rho$ is
$-K_z'/4\pi G$ of the \emph{same} draw, not an independently error-barred
curve.  The point-estimate tower reproduces the vertical-force profile obtained by
Paper~I's independent solver on the same fitted surface to a
root-mean-square $7\times10^{-4}$, so the
machinery is faithful.  What the sister adds is the coherent posterior around
it.

The tower also exposes its own weakest rung.  The force $K_z(1)$ and the surface
density $\Sigma(1)$ cover the truth, but the local density $\rho(0)$ and
vertical frequency $\nu(0)$ (the deepest derivative rung, near the midplane)
sit a little low with a tight band (Figure~\ref{fig:flagship_tower}c).  At the
survey-scale sample of $5\times10^5$ stars the exact-midplane $\rho(0)$ lies
$2.8\,\sigma_{\rm stat}$ below the truth $0.159155$, while with $10^7$ stars the offset
falls to $1.2\,\sigma_{\rm stat}$.  This residual is the derivative-gap
systematic of Paper~I, and the sister's role here is precisely to \emph{separate} it
from the statistical error. A tight statistical posterior around a biased point
is the signature of a discrepancy relative to the \emph{known mock truth}, and
its shrinkage with sample size marks it as a resolvable derivative effect rather
than a fixed floor.  In real data, a narrow conditional posterior would show
only that quoted measurement noise is subdominant under the assumed model, not
identify the culprit systematic.

\subsection{Readable physics, and effective dimension far below $P$}
\label{sec:flagship:physics}

Figures~\ref{fig:flagship_tower}d and \ref{fig:flagship_tower}e carry the
same message as Figures~\ref{fig:rotcurve}d and \ref{fig:rotcurve}e, but
at the scale of our flagship example.  Panel~(d) is the
joint posterior correlation of five genuinely different physical quantities,
three of which differentiate every draw.  The affine $K_z$--$\Sigma$
relation is exactly $-1$, while the nonlinear $\rho$--$\nu$ relation rounds
to $+1$ for this narrow positive posterior. Both are internal consistency
checks.  The density-force cross-correlations ($\rho(0)$ with $K_z(0.5)$ at
$-0.91$, with $K_z(1)$ at $-0.40$) show how measurement noise in the
phase-space fit links the midplane density to the force higher up,
information that is present in the parameter covariance but readable only
here.

Panel~(e) is the practical claim of this paper in one number.  The $K_z$
force posterior, a distribution over an entire function delivered from a fit
with $P=800$ parameters, lives in an effective subspace of dimension
$K_{1\%}=4$. Four posterior deformation modes reproduce $99\%$ of its
variance.  This is an empirical Karhunen--Lo\`eve basis.  The modes and their
cumulative variance are the singular vectors and squared singular values of the
centered ensemble of \emph{re-solved} $K_z$ draws, not of a linearized
transport, because $\vect\beta$ is re-solved per draw
[Eq.~\eqref{eq:cbe_gls}].  The basis is therefore optimal for that empirical
covariance, and unlike the value-space modes of \S\ref{sec:modes} its
coefficients are not necessarily independent standard normals.

The force is expanded in a finite spline basis, so it is worth asking whether
the four-mode count is a property of the posterior or an artifact of that
expansion.  We find that it is the former, as
refining the force basis from $8$ to $32$ cubic
B-splines raises the ambient dimension of the draw ensemble, and hence
its maximum possible rank, from $8$ to $32$ (the measured numerical rank
tracks it exactly), yet leaves $K_{1\%}=4$ and $K_{5\%}=4$ unchanged.
The cumulative variance curves agree to within $2.9\times10^{-3}$
through mode four ($0.603$, $0.835$, $0.932$, $0.997$ at the fiducial
resolution), and the factor $4$ to $6$ gap between the fourth and fifth
singular values persists.  The result is therefore set by the
posterior and not by the resolution of the force representation, though our
test varies only the resolution within the cubic B-spline family, not the
family itself.

Because these are empirical
pushforward coordinates rather than the linear-transport modes of
\S\ref{sec:modes}, a reduced generative model must carry their \emph{joint}
coefficient law.  That law is part of the
declared representation.  This is why a chain over
those $800$ gauge-ridden parameters would spend essentially all of its effort
in directions the chosen force deliverable never sees.  The four-mode count is a
practical statement, the number of coordinates a user must carry to
reproduce this deliverable's posterior, and not a compression ratio
against $P=800$.  The $K_z$ posterior never had $P$ directions available
in any basis, since its ambient dimension is capped by the declared force
representation, and the refinement test above shows that the count does
not grow with that cap.  The result is also local to the
deliverable, applying on the stated $K_z$ grid and metric and not to the
full $g(z,v)$ posterior.

\subsection{Perturbed-refit validation, and the fork made empirical}
\label{sec:flagship:refit}

\begin{figure*}[t]
\centering
\includegraphics[width=\textwidth]{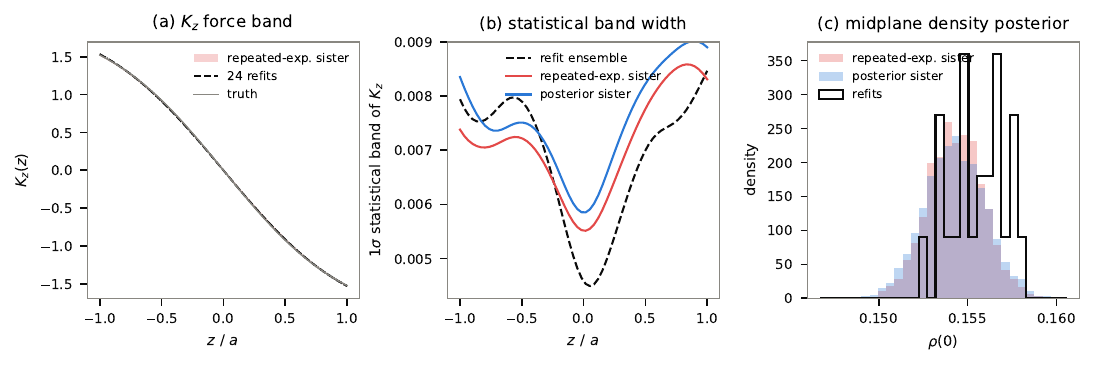}
\caption{The linear-response transport checked against a full nonlinear
  reference for the evidence-selected $\alpha=0.125$ flagship fit
  (\S\ref{sec:flagship:refit}): $24$ independent mock catalogues, each drawn
  fresh at the survey scale ($5\times10^5$ stars) and refit from scratch with the same exact-Poisson
  evidence-regularized procedure on a fresh model instance.  (a)~The
  refit-ensemble $K_z$ band against the repeated-experiment sister band.
  (b)~The $1\sigma$ band-width profiles: refit ensemble, repeated-experiment
  sister, posterior sister.  (c)~The exact $\rho(0)$ distributions: refit
  ensemble (steps), repeated-experiment sister, posterior sister.}
\label{fig:flagship_refit}
\end{figure*}

We next validate the transport against a full nonlinear reference.  Each of $24$
independent mock catalogues with $5\times 10^5$ stars is drawn and refit from
scratch with the same exact-Poisson, evidence-regularized procedure on a fresh
model instance, so the ensemble is the genuine sampling distribution of the
estimator.  Its first-order covariance is the repeated-estimator covariance
$\mat C_{\rm rep}=\mat J_f\mat H^{-1}\mat H_L\mat H^{-1}\mat J_f^\top$, not the
posterior sister $\mat J_f\mat H^{-1}\mat J_f^\top$.  Because the prior matters
here (it is what makes $\mat H$ nonsingular), the two flavors genuinely differ,
and this experiment separates them empirically,
the fork between the two flavors that \S\ref{sec:transport:flavors} could only state for
the flat-prior rotation curve.

Because every catalogue is an independent draw, there is
no centering choice to make and no base-point subtlety, so each is a genuine
repeated experiment.  Every member is fit from the same deterministic seed
initializer, so the branch and path scatter that \S\ref{sec:rotcurve:refit}
exposes by rerunning its initializer is present here as well.  
Only stationarity-certified refits that pass
Eq.~\eqref{eq:stationarity_certificate} enter the ensemble.

Across the core (Figure~\ref{fig:flagship_refit}), the refit-ensemble $K_z$
band width matches the repeated-experiment sister to within the ensemble's
own sampling scatter, the core median standard-deviation ratio being
$0.90\pm0.07$ under member resampling, and the posterior-sister band, wider
by the prior term $\mat J_f\mat H^{-1}\mat P\mat H^{-1}\mat J_f^\top$, sits just
outside it.  The exact-$\rho(0)$ distributions from the three constructions
coincide.  The procedure-matched refits thus show no resolved mismatch with the
repeated-experiment transport at this capacity.

\subsection{Evidence priors and capacity averaging}
\label{sec:flagship:evidence}

A sister at fixed segment count $S$ and prior strength $\alpha$ delivers
$p(M\mid D,S,\alpha)$, not $p(M\mid D)$.  The persistent prior $\mat P$ imported from
the evidence controller is what turns a singular likelihood curvature
into a posterior.  We additionally evaluate a declared grid (the full
Cartesian product) of $S\in\{80,100,120\}$ and
$\alpha\in\{0.03125,0.0625,0.125,0.25,0.5,1,2\}$, with a uniform discrete
prior $\pi_j$ over its 21 members and the Laplace evidence of each
fixed-$\alpha$ fit as its model score.  We call the resulting normalized quantities
\emph{empirical-Bayes Laplace grid weights}, $w_j$, and form
\begin{equation}
\begin{split}
p_{\rm grid}(M\mid D)
  &=\sum_{j=1}^{21}w_jp(M\mid D,S_j,\alpha_j),\\
w_j&\propto p_{\rm Lap}(D\mid S_j,\alpha_j)\pi_j.
\end{split}
\label{eq:capacity_mixture}
\end{equation}
The prior strength $\alpha$ is fixed within each component.  The seed prior anchor and family-RMS
scale template are fixed pre-data functions of topology and seed, as declared
in \S\ref{sec:transport:flavors}. They are not constructed from the observed
counts.  This is nevertheless a finite empirical-Bayes structural audit, not
a claim to integrate over all architectures or hyperparameters.

Every component is independently certified by the stationarity criterion of
\S\ref{sec:rotcurve:cert}.  Four components carry $0.9949$ of the weight:
$(S,\alpha)=(100,0.125)$ dominates at weight
$0.895$ (component mean $0.154797$, $\sigma=0.001679$), followed by
$(100,0.25)$, $(80,0.125)$, and $(100,0.5)$.  The evidence thus selects
$\alpha\simeq0.125$, the value used for the flagship figures.  At exactly $z=0$
the $21$-component mixture has central $68\%$ interval $[0.15289,0.15636]$,
mean $0.154673$, within-component standard deviation $0.001672$, and
between-component standard deviation $0.000382$. Only $5.0\%$ of its total
variance is between components, so the capacity/prior collection adds little
structural uncertainty here.  The analytic truth remains $2.6$ total standard
deviations above the mixture mean, the same residual derivative-gap systematic that
\S\ref{sec:flagship:tower} identified at the scale of the flagship example, and it too shrinks with
sample size.  This remains conditional on the declared model collection.

\section{The Sister as a Reduced Stochastic Representation}
\label{sec:rom}

Thus we see that a mean model plus $K$ deformation modes is more than an uncertainty
report.  It is a reduced stochastic model
\begin{equation}
f(x,\vect\zeta)=\hat f(x)+\sum_{k=1}^{K}s_k\zeta_k\psi_k(x),
\label{eq:fxz}
\end{equation}
a deterministic, differentiable function of a physical coordinate $x$ and an
\emph{uncertainty coordinate} $\vect\zeta\in\reals^K$.  The construction is formally
a reduced-order model.  Unlike a snapshot proper-orthogonal-decomposition
(POD) basis~\citep{Sirovich1987}, an active
subspace~\citep{Constantine2014}, or a likelihood-informed subspace built in
parameter space~\citep{Cui2014}, the sister takes its metric and truncation
from the posterior covariance of a declared function-space deliverable (the
science quantity we want).  Posterior-covariance Karhunen--Lo\`eve expansions, goal-oriented
low-rank approximations~\citep{Spantini2017}, and Bayesian reduced
bases~\citep{Lieberman2010} share parts of this geometry.  Our claim is the
particular end-to-end construction, from bookkeeping of which residual rows
carry noise, through the implicit response of the fit to that noise, to
scientific pushforwards that keep each draw's derivatives coherent, all
checked against direct refits.  Equation~\eqref{eq:mode_truncation} makes
the retained basis optimal for exactly the declared deliverable and
quadrature metric.  Paper~I learns a deterministic scientific function. This
paper learns the geometry of the family of scientifically plausible
functions around it.  Together they are the representation.

\section{Conclusions}
\label{sec:conclusions}


This paper represents the conditional local posterior of a scientific
surrogate as a low-dimensional basis of coherent function deformations:
differentiate the fitted stationary branch, compress the resulting
measurement-to-function transport by its singular functions, and push each
coherent draw through the scientific calculation.  On the rotation-curve
suite, the construction agrees with an independent spline posterior.
Stationary warm refits reproduce fixed-branch function widths at the
10--15\% level, while cold starts reveal additional branch and path scatter.
In the $P=800$ flagship problem, four modes carry
99\% of the $K_z$ posterior variance, and the evidence-selected prior leaves a
small residual midplane-density systematic, $2.8\,\sigma$ at $5\times 10^5$ stars
and $1.2\,\sigma$ at $10^7$ stars, which shows a resolvable derivative-gap effect, not a
fixed floor.

The scope is deliberately local.  The construction is first order on one
smooth branch and is conditional on the declared model capacity,
empirical-Bayes strengths, active sets, and robust weights.  It does not
represent disconnected basins, model discrepancy, or non-Gaussian response
along weakly constrained directions. The cold-refit, cutoff, and $Q$
audits expose those boundaries.

The construction as written in Eq.~\eqref{eq:sister} contains two
assumptions that are easy to conflate, that the local posterior over
functions is Gaussian and that it is low dimensional.  Only the second is needed for the results reported
here.  The first is
already partly relaxed, since nonlinear compressions are pushed exactly rather
than linearized (\S\ref{sec:compressions}), the flagship problem likelihood is an exact
Poisson deviance, and the re-solved force modes of
\S\ref{sec:flagship:physics} carry an empirical joint coefficient law that is
not a product of standard normals.  What currently remains Gaussian is the coefficient
law of the value-space modes in Eq.~\eqref{eq:sister}, and replacing it by a
non-Gaussian law over the same $K$ coordinates would leave the mode basis, the
compressions, and every deliverable pushforward unchanged.

The intended endpoint is that, after fitting, NestyNet emits the sister
in its portable form,
\begin{center}
mean model $+$ $K$ deformation modes $+$ a law for $K$ coefficients,
\end{center}
so a user can evaluate scientific deliverables over coherent function draws
without interacting with the network parameterization.  Paper~I exposed the
scientifically meaningful fitted function. This paper exposes the geometry
of the family of functions permitted by the measurements.

There is a larger reason to want scientific knowledge in this form.  The
user who evaluates deliverables over coherent draws need not be human.  A
fitted model that carries its own uncertainty coordinates, with a declared
coefficient law and a certificate of convergence, is a piece of knowledge
that an automated system can query, propagate through a calculation, and
be corrected by, all without touching the network parameterization.
Later papers in this series attach the sister to learned world models, so that a
representation can state in calibrated terms how far its own beliefs
reach.  The sister is the unit in which a machine can hold what a
measurement taught it, and how well.

\begin{acknowledgments}
RI gratefully acknowledges funding in the initial stages of this project
from the European Research Council (ERC) under the European Union's Horizon
2020 research and innovation programme (grant agreement No. 834148). We
gratefully acknowledge the High Performance Computing center of the
Universit\'e de Strasbourg for a very generous time allocation and for
their support over the development of this project.
\end{acknowledgments}

\software{
NestyNet\_stat (this work;
\url{https://github.com/RodrigoIbata/NestyNet_stat}), is based on
NestyNet \citep{NestyNet2026a},
PyTorch \citep{Paszke2019},
NumPy \citep{Harris2020},
SciPy \citep{Virtanen2020},
Matplotlib \citep{Hunter2007}.}

\bibliographystyle{aasjournalv7}
\bibliography{nestynet_paper2}

\end{document}